\documentclass[%
 reprint,
 amsmath,amssymb,
 aps,
]{revtex4-1}

\usepackage{graphicx}
\usepackage{dcolumn}
\usepackage{bm}

\begin{document}

\preprint{APS/123-QED}

\title{Detecting quantum correlations by means of local noncommutativity\\}

\author{Yu Guo}
\altaffiliation{Email address: guoyu3@yahoo.com.cn}
\affiliation{Department of Mathematics, Taiyuan University of Technology, Taiyuan 030024, China}
\affiliation{Department of Mathematics, Shanxi Datong University, Datong 037009, China}
\affiliation{Department of Physics and Optoelectroincs, Taiyuan University of Technology, Taiyuan 030024, China}
\author{Jinchuan Hou}
\altaffiliation{Email address: jinchuanhou@yahoo.com.cn}
\affiliation{{Department of Mathematics, Taiyuan University of Technology, Taiyuan 030024, China}}

\date{ \today}

\begin{abstract}
Quantum correlation is a key to our understanding of quantum
physics. In particular, it is essential for the powerful
applications to quantum information and quantum computation. There
exist quantum correlations beyond entanglement, such as quantum
discord (QD) and measurement-induced nonlocality (MiN) [Phys. Rev.
Lett. \textbf{106}, 120401(2011)]. In [Phys. Rev. A \textbf{77},
022113(2008)], a subclass of PPT states so-called strong positive
partial transposition (SPPT) states was introduced and it was
conjectured there that SPPT states are separable. However, it was
illustrated with examples in [Phys. Rev. A \textbf{81},
064101(2010)] that this conjecture is not true. Viewing the original
SPPT as SPPT up to part B, in the present paper, we define SPPT
state up to part A and B respectively and present a separable class
of SPPT states, that is the super SPPT (SSPPT) states, in terms of
local commutativity. In addition, classical-quantum (CQ) states and
nullity of MiN are characterized via local commutativity. Based on
CQ states, the geometric measure of quantum discord (GMQD) for
infinite-dimensional case is proposed. Consequently, we highlight
the relation among MiN, QD(GMQD), SSPPT and separability through a
unified approach for both finite- and infinite-dimensional systems:
zero MiN implies zero QD(GMQD), zero QD(GMQD) signals SSPPT and
SSPPT guarantees separability, but the converses  are not valid.
\begin{description}
\item[PACS numbers]03.65.Ud, 03.65.Db, 03.67.Mn.
\end{description}
\end{abstract}

\pacs{Valid PACS appear here}
\maketitle


\section{Introduction}

Correlations among subsystems of a composite quantum system, with
fundamental applications in many typical of the fields of quantum
information and quantum computing
\cite{Nielsen,Horodecki,Guhne,Ollivier,Henderson,Dakic,
Knill,Datta1,Datta2,Datta3,Lanyon,Qasimi,Batle,Luo1,Guo5}, are
usually studied in entanglement-verses-separability framework.
However, apart from entanglement, quantum states display other
quantum correlations, such as quantum discord (QD)
\cite{Ollivier,Henderson} and measurement-induced nonlocality (MiN)
\cite{Luo1}. Entanglement lies at the heart of quantum information
theory \cite{Horodecki,Guhne}. QD can be used in quantum computation
\cite{Dakic,Qasimi}. It is indicated in \cite{Luo1} that MiN may
also be applied in quantum cryptography, general quantum dense
coding \cite{Mattle,Li}, remote state control \cite{Bennett,Peters},
etc.. In particular, it is of primary importance to test whether a
given quantum state has quantum correlation in it.

Consider the two-mode system labeled by A+B which is described by a
complex Hilbert space $H=H_A\otimes H_B$ with $\dim H_A\otimes
H_B\leq+\infty$. We denote by $\mathcal{S}(H_A\otimes H_B)$ the set
of all states acting on $H_A\otimes H_B$, that is, the set of all
positive operators with trace one in ${\mathcal T}(H_A\otimes H_B)$,
the space of all trace-class operators. By definition, a quantum
state $\rho\in{\mathcal S}(H_A\otimes H_B)$ is \emph{separable} if
it can be written as
\begin{eqnarray}
\rho=\sum_ip_i\rho_i^A\otimes\rho_i^B,\quad \sum_ip_i=1,~
p_i\geq0 \label{d}
\end{eqnarray}
or can be approximated in the trace norm by the states of the above
form \cite{Werner,HP2}. Otherwise, $\rho$ is called
\emph{entangled}. In particular, a separable state with the form as
in Eq.(\ref{d}) is called \emph{countably separable}
\cite{Holevo,Guo3}. If $\dim H_A\otimes H_B<+\infty$, then all
separable states $\rho$ acting on $H_A\otimes H_B$ are countably
separable \cite{HP2}. But, in the infinite-dimensional case, there
exists separable states which are not countably separable
\cite{Holevo}.

One of the most famous criteria for detecting entanglement is the
so-called positive partial transpose (PPT) criterion proposed by
Peres and Horodecki \cite{Peres,Horodecki2}: if a quantum state
$\rho$ acting on the Hilbert space $H_A\otimes H_B$ is separable,
then the partial transposes of $\rho$ are positive operators, that
is, $\rho^{T_{A/B}}\geq0$. There exist entangled PPT states (i.e.,
states with positive partial transpose) except for $2\otimes 2$ and
$2\otimes 3$ systems \cite{Horodecki2}. Consequently, it is
important to know which PPT states are separable and which are
entangled (PPT entangled states are known as bound states which are
not  distillable). In \cite{Chruscinski}, a subclass of PPT states,
called \emph{strong positive partial transpose} (SPPT) states, is
considered according to the Cholesky decomposition of positive
semidefinite matrix. These states have ``strong PPT property'' which
is insured by the canonical construction (see below). Based on
several examples of SPPT states, it is conjectured in
\cite{Chruscinski} that all SPPT states are separable. However, it
is not true since there exist entangled states which are SPPT
\cite{Ha2}. SPPT can be used for witnessing quantum discord in
$2\otimes n$ systems \if false (see Theorem 1 in \cite{Bylicka}):
any classical-quantum (CQ) state in $m\otimes n$ system is SPPT \fi
because a state $\rho$ in finite-dimensional bipartite system
associated with $H_A\otimes H_B$ is CQ if and only if it has zero QD
with respect to part A \cite{Ollivier}. In this letter, we propose a
special class of SPPT states which we call super SPPT (SSPPT)
states. It turns out every SSPPT state is countably separable and
any CQ (resp. QC) state is SSPPT up to part B (resp. A).
Furthermore, we find that SSPPT up to part A/B can detect QD with
respect to part A/B, and there exist zero QD states with nonzero
MiN.

The quantum discord can be viewed as a measure of the minimal loss
of correlation in the sense of quantum mutual information. Recall
that the quantum discord of a state $\rho$ on finite-dimensional
Hilbert space $H_A\otimes H_B$ is defined in \cite{Ollivier} by
\begin{eqnarray}
D_A(\rho):=\min_{\Pi^A}\{I(\rho)-I(\rho|\Pi^A)\},
\end{eqnarray}
where, the minimum is taken over all local von Neumann measurements
$\Pi^A$,
\begin{eqnarray*}
I(\rho):=S(\rho_A)+S(\rho_B)-S(\rho)
\end{eqnarray*}
is interpreted as the quantum mutual information,
\begin{eqnarray*}
S(\rho):=-{\rm Tr}(\rho\log\rho)
\end{eqnarray*}
is the von Neumann entropy,
\begin{eqnarray*}
&I(\rho|\Pi^A)\}:=S(\rho_B)-S(\rho|\Pi^A),&\\
&S(\rho|\Pi^A):=\sum_kp_kS(\rho_k),&
\end{eqnarray*}
and
\begin{eqnarray*}
\rho_k=\frac{1}{p_k}(\Pi_k^A\otimes I_B)\rho(\Pi_k^A\otimes I_B)
\end{eqnarray*}
with $p_k={\rm Tr}[(\Pi_k^A\otimes I_B)\rho(\Pi_k^A\otimes I_B)]$,
$k=1$, 2, $\dots$, $\dim H_A$. Throughout this paper, all logarithms
are taken to base 2. QD of any state is nonnegative
\cite{Ollivier,Datta}. Also recall that a state $\rho$ on
$H_A\otimes H_B$ is said to be a CQ state if it has the form of
\begin{eqnarray}
\rho=\sum_ip_i|i\rangle\langle i|\otimes\rho_i^B,\label{y}
\end{eqnarray}
for some orthonormal set $\{|i\rangle\}$ of $H_A$, where $\rho_i^B$s
are states of the subsystem B, $p_i\geq0$, $\sum_ip_i=1$. It is
known that a state has zero QD if and only if it is a CQ state. The
conditions for nullity of quantum discord may be found in
\cite{Bylicka,Datta,Dakic}.

Measurement-induced nonlocality was firstly proposed
by Luo and Fu \cite{Luo1}, which can be viewed as a
kind of quantum correlation from a geometric perspective
based on the local von Neumann measurements
from which one of the reduced states is left invariant.
The MiN of $\rho$, denoted by $N_A(\rho)$, is
defined in \cite{Luo1} by
\begin{eqnarray}
N_A(\rho):=\max_{\Pi^A} \|\rho-\Pi^A(\rho)\|_2^2,
\end{eqnarray}
where $\|\cdot\|_2$ stands for the Hilbert-Schmidt norm (that is
$\|A\|_2=[{\rm Tr}(A^\dag A)]^{\frac{1}{2}}$), and the maximum is
taken over all local von Neumann measurement $\Pi^A=\{\Pi_k^A\}$
with
\begin{eqnarray*}
&\sum_k\Pi_k^A\rho_A\Pi_k^A=\rho_A,&\\
&\Pi^A(\rho)=\sum_k(\Pi_k^A\otimes I_B)\rho(\Pi_k^A\otimes I_B).&
\end{eqnarray*}
MiN is different from, and in some sense dual to,
the \emph{geometric measure of quantum discord}(GMQD) \cite{Luo3}
\begin{eqnarray*}
D_G(\rho):=\min_{\Pi^A}\|\rho-\Pi^A(\rho)\|_2^2
\end{eqnarray*}
where
$\Pi^A$ runs over \emph{all} local von Neumann measurements
(GMQD is originally introduced in \cite{Dakic} as
\begin{eqnarray*}
D_G(\rho):=\min_{\chi}\|\rho-\chi\|_2^2
\end{eqnarray*}
with $\chi$ runs over all zero QD states. It is proved in
\cite{Luo3} that the two definitions are equivalent).

Mathematically, quantumness is always associated with
non-commutativity while classical mechanics displays commutativity
in some sense \cite{Bastos,Douglas,Gamboa}. With this idea in mind,
it is possible to describe these quantum correlations mentioned
above in terms of non-commutativity. The aim of this paper is to
find a unified mathematical language in these quantum correlations
analysis, and hence shine new light on the structure of quantum
correlation, from which we can get more understanding of these
different kinds of quantum correlations.

The remainder of this paper is organized as follows. In Sec.II, we
define SSPPT states for both finite- and infinite-dimensional
systems and prove that these states are separable. Then,  in
Sec.III, CQ states are characterized, from which we show that CQ is
equivalent to zero GMQD and then SSPPT can detect QD, GMQD. Sec.IV
distributes to giving a necessary and sufficient condition for a
state to have zero MiN. A summary is given in the last section.

\section{Super strong positive partial transpose states}

In this section we first give definitions of
SSPPT states for both finite-
and infinite-dimensional systems.
Then we show that all SSPPT states are separable.

\subsection{Definitions}

\emph{Finite-dimensional case}\ \  In a $m\otimes n$ system with
$mn<+\infty$, any state $\rho$ may be viewed as a block $m\times
m$ matrix with $n\times n$ blocks. Due to the Cholesky
decomposition, there exists block upper triangular matrix $X$
($m\times m$ block  matrix with $n\times n$ blocks),
\begin{eqnarray*}
X=\left(\begin{array}{c|c|c|c|c}
X_1&S_{12}X_1&S_{13}X_1&\cdots&S_{1m}X_1 \\ \hline
0&X_2&S_{23}X_2&\cdots&S_{2m}X_2 \\ \hline
\vdots&\vdots&\ddots&\vdots&\vdots\\ \hline
0&0&0&X_{m-1}&S_{m-1,m}X_{m-1}\\ \hline
0&0&0&0&X_m\end{array}\right)
\end{eqnarray*}
such that $\rho=X^\dag X$ (the choice of $X$ is not unique). If
$\rho^{T_A}=Y^\dag Y$ with
\begin{eqnarray*}
Y=\left(\begin{array}{c|c|c|c|c}
X_1&S_{12}^\dag X_1&S_{13}^\dag X_1&\cdots&S_{1m}^\dag X_1 \\ \hline
0&X_2&S_{23}^\dag X_2&\cdots&S_{2m}^\dag X_2 \\ \hline
\vdots&\vdots&\ddots&\vdots&\vdots\\ \hline
0&0&0&X_{m-1}&S_{m-1,m}^\dag X_{m-1}\\ \hline
0&0&0&0&X_m\end{array}\right),
\end{eqnarray*} $\rho$ is called a SPPT state \cite{Chruscinski}.
Obviously, if $S_{ij}$s satisfy the condition
\begin{eqnarray}
[S_{ki},S_{kj}^\dag]=0,\quad k<i\leq j,\label{u}
\end{eqnarray}
then $\rho$ must be SPPT \cite{Chruscinski}. In such a case, we
say that $\rho$ is a super SPPT (or SSPPT briefly) state.

Let $\{|i\rangle\}$ and $\{|j'\rangle\}$ be
the canonical computational bases of
$\mathbb{C}^m$ and $\mathbb{C}^n$, respectively.
Then $\rho$ can be presented as
\begin{eqnarray}
\rho=\sum_{i,j}A_{ij}\otimes F_{ij}=\sum_{k,l}E_{kl}\otimes B_{kl},
\end{eqnarray}
where $F_{ij}=|i'\rangle\langle j'|$ and $E_{kl}=|k\rangle\langle
l|$. That is, a state $\rho$ acting on $H_A\otimes H_B$ can be
represented as
\begin{eqnarray}
\rho=[B_{kl}] \text{ or }{\rho}=[A_{ij}].\label{t}
\end{eqnarray}
Symmetrically, we can define SPPT and SSPPT states up to part A. Namely,
if there exist
\begin{eqnarray*}
\tilde{X}=\left(\begin{array}{c|c|c|c|c}
\tilde{X}_1&\tilde{S}_{12}\tilde{X}_1&\tilde{S}_{13}\tilde{X}_1
&\cdots&\tilde{S}_{1m}\tilde{X}_1 \\ \hline
0&\tilde{X}_2&\tilde{S}_{23}\tilde{X}_2
&\cdots&\tilde{S}_{2m}\tilde{X}_2 \\ \hline
\vdots&\vdots&\ddots&\vdots&\vdots\\ \hline
0&0&0&\tilde{X}_{m-1}&\tilde{S}_{m-1,m}\tilde{X}_{m-1}\\ \hline
0&0&0&0&\tilde{X}_m\end{array}\right)
\end{eqnarray*}
and
\begin{eqnarray*}
\tilde{Y}=\left(\begin{array}{c|c|c|c|c}
\tilde{X}_1&\tilde{S}_{12}^\dag\tilde{X}_1
&\tilde{S}_{13}^\dag\tilde{X}_1&\cdots
&\tilde{S}_{1m}^\dag\tilde{X}_1 \\ \hline
0&\tilde{X}_2&\tilde{S}_{23}^\dag\tilde{X}_2
&\cdots&\tilde{S}_{2m}^\dag\tilde{X}_2 \\ \hline
\vdots&\vdots&\ddots&\vdots&\vdots\\ \hline
0&0&0&\tilde{X}_{m-1}&\tilde{S}_{m-1,m}^\dag\tilde{X}_{m-1}\\ \hline
0&0&0&0&\tilde{X}_m\end{array}\right)
\end{eqnarray*}
such that
\begin{eqnarray*}
{\rho}=[A_{ij}]=\tilde{X}^\dag \tilde{X}
\end{eqnarray*}
and
\begin{eqnarray*}
{\rho}^{T_B}=\tilde{Y}^\dag \tilde{Y},
\end{eqnarray*}
then we call $\rho$ is SPPT up to part A. In particular, if
$\tilde{S}_{ij}$s satisfy
\begin{eqnarray*}
[\tilde{S}_{ki},\tilde{S}_{kj}^\dag]=0,\quad k<i\leq j,\label{v}
\end{eqnarray*}
we call that $\rho$ is SSPPT up to part A.

\emph{Infinite-dimensional case}\ \  Note that the Cholesky
factorization can be generalized to (not necessarily finite)
matrices with operator entries, so we can define SPPT and SSPPT
states for infinite-dimensional bipartite systems.

Assume that $\dim H_A\otimes H_B=+\infty$, $\dim H_A=+\infty$,
$\{|i\rangle\}$ and $\{|j'\rangle\}$ be any orthonormal bases of
$H_A$ and $H_B$, respectively. Let $E_{kl}=|k\rangle\langle l|$.
Consequently, any state $\rho$ acting on $H_A\otimes H_B$ can be
represented by
\begin{eqnarray}
\rho=\sum_{k,l}E_{kl}\otimes B_{kl},\label{a}
\end{eqnarray}
where $B_{kl}$s are trace class operators
on $H_B$ and the series converges in
trace norm \cite{Guo}.

That is,
\begin{eqnarray}
\rho=\left(\begin{array}
{c|c|c|c|c}
B_{11}&B_{12}&B_{13}&\cdots&\cdots \\ \hline
B_{21}&B_{22}&B_{23}&\cdots&\cdots \\ \hline
\vdots&\vdots&\ddots&\vdots&\cdots\\ \hline
B_{n1}&B_{n2}&B_{n3}&\cdots&\cdots\\ \hline
\vdots&\vdots&\vdots&\cdots&\ddots
\end{array}\right)\label{w}
\end{eqnarray}
under the given bases. With respect to these bases, if there exist
some upper triangular (infinite) operator matrices of the form
\begin{eqnarray*}
X=\left(\begin{array}{c|c|c|c|c|c}
X_1&S_{12}X_1&S_{13}X_1&\cdots&S_{1m}X_1&\cdots \\ \hline
0&X_2&S_{23}X_2&\cdots&S_{2m}X_2&\cdots \\ \hline
\vdots&\vdots&\ddots&\vdots&\vdots&\cdots\\ \hline
0&0&0&X_{m-1}&S_{m-1,m}X_{m-1}&\cdots\\ \hline
0&0&0&0&X_m&\cdots\\ \hline
0&0&0&0&0&\ddots
\end{array}\right)
\end{eqnarray*}
and
\begin{eqnarray*}
Y=\left(\begin{array}{c|c|c|c|c|c}
X_1&S_{12}^\dag X_1&S_{13}^\dag X_1
&\cdots&S_{1m}^\dag X_1&\cdots \\ \hline
0&X_2&S_{23}^\dag X_2&\cdots&S_{2m}^\dag X_2&\cdots \\ \hline
\vdots&\vdots&\ddots&\vdots&\vdots&\cdots\\ \hline
0&0&0&X_{m-1}&S_{m-1,m}^\dag X_{m-1}&\cdots\\ \hline
0&0&0&0&X_m&\cdots\\ \hline
0&0&0&0&0&\ddots\end{array}\right)
\end{eqnarray*}
with the same size as that of $\rho=(B_{kl})$ in Eq.(\ref{w}) such
that
\begin{eqnarray*}
\rho=X^\dag X
\end{eqnarray*}
\text{ and }
\begin{eqnarray*}
 \rho^{T_A}=Y^\dag Y,
\end{eqnarray*}
then we call that $\rho$ is a SPPT state up to part B. Moreover, if
$S_{ij}$s are diagonalizable  normal operators and satisfy
\begin{eqnarray}
[S_{ki},S_{kj}^\dag]=0,\quad k<i\leq j,
\end{eqnarray}
we call that $\rho$ is a SSPPT state up to part B. Note that, $X$
is a Hilbert-Schmidt operator, and in the case that $\rho$ is
SSPPT, $S_{ij}$ is normal for any $k<i\leq j$.

Symmetrically, any state $\rho$ acting on
$H_A\otimes H_B$ can be represented by
\begin{eqnarray}
\rho=\sum_{i,j}A_{ij}\otimes F_{ij},\label{z}
\end{eqnarray}
where $F_{ij}=|i'\rangle\langle j'|$, $A_{ij}$s are trace class
operators on $H_A$ and the series converges in trace norm
\cite{Guo}. Analogy to the finite-dimensional case, we can define
SSPPT state up to part A when regarding $\rho$ as
${\rho}=[A_{ij}]$. Furthermore, it is worth mentioning that SSPPT
up to A is not equivalent to that up to B.

{\sl Remark.}\quad It is easily checked that the SPPT and the SSPPT
are invariant under the local unitary operation. So, the definitions
of SPPT and SSPPT are independent to the choice of local bases
$\{|i\rangle\}$ and $\{|j'\rangle\}$ of $H_A$ and $H_B$
respectively. Namely, if $\rho$ is SPPT (or SSPPT) with respect to
the given local bases $\{|i\rangle\}$ and $\{|j'\rangle\}$ of $H_A$
and $H_B$ respectively, then it is also SPPT (or SSPPT) with respect
to the other choice of local bases.

\subsection{SSPPT states are separable}

The main result of this section is the following.

{\sl Theorem 1.}\quad Let $\rho\in\mathcal{S}(H_A\otimes H_B)$
with $\dim H_A\otimes H_B\leq+\infty$ be a SSPPT state up to part
A or B. Then $\rho$ is countably separable.

{\sl Proof.}\quad We only need to check the case of SSPPT up to
part B since the proof  for the case of SSPPT up to A is similar.
We consider the infinite-dimensional case, the finite-dimensional
is then obvious.

Let $\rho$ be a SSPPT state up to part B. Then $\rho=X^\dag X$ and
$\rho^{T_A}=Y^\dag Y$, where $X$ and $Y$ are upper triangular
operator matrices of the form mentioned above. Let $C_k$ be the
infinite operator matrix with the same size as that of $X$, which
is induced from $X$ by replacing all entries by zero except for
the $k$th row of $X$, i.e.,
\begin{widetext}
\begin{eqnarray*}
C_k=\left(\begin{array}{c|c|c|c|c|c|c|c|c}
0&\cdots&0&0&0&0&\cdots&0&\cdots \\ \hline
\vdots&\ddots&\vdots&\vdots&\vdots
&\vdots&\vdots&0&\cdots \\ \hline
0&\cdots&0&0&0&0&\cdots&0&\cdots \\ \hline
0&\cdots&0&X_k&S_{k,k+1}X_k&S_{k,k+2}X_k
&\cdots&S_{km}X_k&\cdots \\ \hline
0&\cdots&0&0&0&0&\cdots&0&\cdots \\ \hline
\vdots&\cdots&\vdots&\vdots&\vdots&\vdots
&\cdots&\vdots&\ddots\end{array}\right),\quad k=1, 2, \dots.
\end{eqnarray*}
\end{widetext}
Then $C_i$ is a Hilbert-Schmidt operator and
\begin{eqnarray}
\rho=\sum_{i}C_i^\dag C_i, \quad C_i^\dag C_i\geq0.
\end{eqnarray}
Write $C_i^\dag C_i=p_i\rho_i$ where $ p_i={\rm Tr}(C_i^\dag
C_i)$. We have
\begin{widetext}
\begin{eqnarray}
p_1\rho_1=\left(\begin{array}{c|c|c|c|c|c}
X_1^\dag X_1&X_1^\dag S_{12}X_1&X_1^\dag S_{13}X_1
&\cdots&X_1^\dag S_{1m}X_1 &\cdots\\ \hline
X_1^\dag S_{12}^\dag X_1&X_1^\dag S_{12}^\dag S_{12}X_1
&X_1^\dag S_{12}^\dag S_{13}X_1&\cdots
&X_1^\dag S_{12}^\dag S_{1m}X_1 &\cdots\\ \hline
\vdots&\vdots&\vdots&\vdots&\vdots &\vdots\\ \hline
X_1^\dag S_{1,m-1}^\dag X_1&X_1^\dag S_{1,m-1}^\dag S_{12}X_1
&X_1^\dag S_{1,m-1}^\dag S_{13}X_1&\cdots
&X_1^\dag S_{1,m-1}^\dag S_{1m}X_1&\cdots\\ \hline
X_1^\dag S_{1m}^\dag X_1&X_1^\dag S_{1m}^\dag S_{12}X_1
&X_1^\dag S_{1m}^\dag S_{13}X_1&\cdots
&X_1^\dag S_{1m}^\dag S_{1m}X_1&\cdots\\ \hline
\vdots&\vdots&\vdots&\cdots&\vdots&\ddots
\end{array}\right).
\end{eqnarray}
\end{widetext}
We claim that $\rho_1$ is a countably separable state. Note that
$X_1$ is a Hilbert-Schmidt operator and  $S_{1l}$s are mutually
commuting diagonalizable normal operators on $H_B$. Thus
\begin{eqnarray*}
&X_1^\dag X_1=\sum_ia_i|\psi_i\rangle\langle\psi_i|&
\end{eqnarray*}
and
\begin{eqnarray*}
&S_{1l}=\sum_jb_j^{(l)}|\phi_j\rangle\langle\phi_j|,\quad l=2, 3,
\dots&
\end{eqnarray*}
for some orthonormal bases $\{|\psi_i\rangle\}$ and
$\{|\phi_j\rangle\}$   of $H_B$. Denote
\begin{eqnarray*}
\beta_{ij}=\langle\psi_i|\phi_j\rangle,
\end{eqnarray*}
\begin{widetext}
\begin{eqnarray}A_i=a_i\left(\begin{array}{cccccc}
1&\sum_jb_j^{(2)}|\beta_{ij}|^2&\sum_jb_j^{(3)}|\beta_{ij}|^2
&\cdots&\sum_jb_j^{(m)}|\beta_{ij}|^2&\cdots \\
\sum_j\bar{b}_j^{(2)}|\beta_{ij}|^2&\sum_j|b_j^{(2)}|^2|\beta_{ij}|^2
&\sum_j\bar{b}_j^{(2)}b_j^{(3)}|\beta_{ij}|^2&\cdots
&\sum_j\bar{b}_j^{(2)}b_j^{(m)}|\beta_{ij}|^2&\cdots \\
\vdots&\vdots&\vdots&\vdots&\vdots&\vdots\\
\sum_j\bar{b}_j^{(m-1)}|\beta_{ij}|^2
&\sum_j\bar{b}_j^{(m-1)}b_j^{(2)}|\beta_{ij}|^2
&\sum_j\bar{b}_j^{(m-1)}b_j^{(3)}|\beta_{ij}|^2
&\cdots&\sum_j\bar{b}_j^{(m-1)}b_j^{(m)}|\beta_{ij}|^2&\cdots\\
\sum_j\bar{b}_j^{(m)}|\beta_{ij}|^2
&\sum_j\bar{b}_j^{(m)}b_j^{(2)}|\beta_{ij}|^2
&\sum_j\bar{b}_j^{(m)}b_j^{(3)}|\beta_{ij}|^2
&\cdots&\sum_j|b_j^{(m)}|^2|\beta_{ij}|^2&\cdots\\
\vdots&\vdots&\vdots&\cdots&\vdots&\ddots
\end{array}\right)
\end{eqnarray}
\end{widetext}
and
\begin{eqnarray*}
B_i=|\psi_i\rangle\langle\psi_i|.
\end{eqnarray*}
Then we have
\begin{eqnarray}
p_1\rho_1=\sum_iA_i\otimes B_i.
\end{eqnarray}
So $A_i\geq 0$ is a trace-class operator for each $i$ and
$\sum_i{\rm Tr}(A_i)=p_1\leq 1$.
\if false Let
\begin{eqnarray*}
|\beta_{j}\rangle=
\left(\begin{array}{c}
1\\
\bar{b}_j^{(2)}\\
\bar{b}_j^{(3)}\\
\vdots\\
\bar{b}_j^{(m)}\\
\vdots
\end{array}\right),\quad j=1, 2, \dots
\end{eqnarray*}
Then
\begin{eqnarray*}
A_i=a_i\sum_j|\beta_{ij}|^2|\beta_j\rangle\langle\beta_j|,\quad i=1,
2, \dots
\end{eqnarray*}
as $\sum_j|\beta_{ij}|^2=\||\psi_i\rangle\|=1$. It follows that
$$ \sum _l|b_j^{(l)}|^2={\rm
Tr}(|\beta_j\rangle\langle\beta_j|)<\infty
$$
whenever $\beta_{ij}\not= 0$, where $b_j^{(1)}=1$. Thus
$|\beta_j\rangle$ is a vector in $H_A$ if  $\beta_{ij}\not= 0$.\fi
Now it is clear that  $\rho_1$ is countably separable.

Similarly, $\rho_i$ is countably separable for each $i$, $i\geq1$.
Hence, $\rho$ is a countably separable state.\hfill$\blacksquare$

 In some sense, diagonalizability can be regarded as a kind of
commutativity since $A$ is normal  implies  $[A,A^\dag]=0$. Thus
Theorem 1 indicates that diagonalizability of $S_{ij}$s and
commutativity between $S_{ij}$s guarantee the separability of
$\rho$.

Obviously, SSPPT is not a necessary condition of separability. In
fact, for SSPPT states $\rho_1$ and $\rho_2$, their convex
combination $t\rho_1+(1-t)\rho_2$ may not be a SSPPT state in
general. That is, the set consisting of SSPPT states is not convex,
while both the set of PPT states and the set of separable states are
convex. However, for pure states, SPPT, SSPPT and separability are
equivalent since a pure state is separable if and only if it is PPT
\cite{GY}. A little more can be said. In fact we have  the following
conclusion.

\textit{Proposition 1.}\quad  Every product state is SSPPT, and a
pure state is separable if and only if it is SPPT (or SSPPT).

By Theorem 1, we obtain some simple separability criteria for
states in $2\otimes n$ (resp. $n\otimes 2$) systems.

\textit{Corollary 1.}\quad Assume that $\dim H_A\otimes
H_B<+\infty$ and $\dim H_A=2$ (or $\dim H_B=2$),
$\rho\in\mathcal{S}(H_A\otimes H_B)$. Write
\begin{eqnarray*}
\rho=\left(\begin{array}{cc}
\rho_{11}&\rho_{12}\\
\rho_{21}&\rho_{22}\end{array}\right) \text{ (or }{\rho}=
\left(\begin{array}{cc}\tilde{\rho}_{11}&\tilde{\rho}_{12}\\
\tilde{\rho}_{21}&\tilde{\rho}_{22}\end{array}\right))
\end{eqnarray*}
as in Eq.(\ref{t}). Then the following statements are true:

(i) If $\rho$ is a SPPT sate up to part B (or, up to part A) and
$\rho_{11}$ (or $\tilde{\rho}_{11}$) is invertible, then $\rho$ is
separable.

(ii) If $\rho_{11}\geq\rho_{22}$ or $\rho_{22}\geq\rho_{11}$ (or,
$\tilde{\rho}_{11}\geq\tilde{\rho}_{22}$ or
$\tilde{\rho}_{22}\geq\tilde{\rho}_{11}$), Then $\rho$ is
separable.

\textit{Proof.}\ \ We only give a proof of (ii) here. No loss of
generality, we assume that $\dim H_A=2$, $
\rho=\left(\begin{array}{cc}
\rho_{11}&\rho_{12}\\
\rho_{21}&\rho_{22}\end{array}\right)$ with $\rho_{ij}\in{\mathcal
T}(H_B)$ and $\rho _{22}\leq \rho_{11}$. We shall show that $\rho$
is SSPPT and hence is separable by Theorem 1.

Since  $\rho\geq 0$ and $\rho_{22}\leq \rho_{11}$, there are
contractive operators $T, S\in{\mathcal B}(H_B)$ with $\ker
T\cap\ker S\cap\ker S^\dag\supseteq \ker \rho_{11}$ such that
$\rho_{12}=\sqrt{\rho_{11}}T\sqrt{\rho_{22}}$ and
$\sqrt{\rho_{22}}=\sqrt{\rho_{11}}S=S^\dag \sqrt{\rho_{11}}$. Let
$S_{12}=TS^\dag$. Then we have
$\rho_{12}=\sqrt{\rho_{11}}S_{12}\sqrt{\rho_{11}}$. Note that
$$\sqrt{\rho_{11}}S_{12}^\dag S_{12}\sqrt{\rho_{11}}=\sqrt{\rho_{22}}T^\dag
T\sqrt{\rho_{22}}\leq \rho_{22}.
$$
Let $$X_2=[\rho_{22}-\sqrt{\rho_{11}}S_{12}^\dag
S_{12}\sqrt{\rho_{11}}]^{\frac{1}{2}},$$
$$X=\left(\begin{array}{cc} \sqrt{\rho_{11}}
&S_{12}\sqrt{\rho_{11}}\\ 0& X_2\end{array}\right)
$$
and
$$Y=\left(\begin{array}{cc} \sqrt{\rho_{11}}
&S_{12}^\dag\sqrt{\rho_{11}}\\ 0& X_2\end{array}\right).
$$
Then $$\begin{array}{rl}\rho=&X^\dag X\\=&\left(\begin{array}{cc}
{\rho_{11}}
&\sqrt{\rho_{11}}S_{12}\sqrt{\rho_{11}}\\
\sqrt{\rho_{11}}S_{12}^\dag\sqrt{\rho_{11}}&
\sqrt{\rho_{11}}S_{12}^\dag S_{12}\sqrt{\rho_{11}}+X_2^\dag
X_2\end{array}\right)\end{array}$$ and
$$\begin{array}{rl}\rho ^{T_A}=&Y^\dag Y\\=&\left(\begin{array}{cc}
{\rho_{11}}
&\sqrt{\rho_{11}}S_{12}^\dag\sqrt{\rho_{11}}\\
\sqrt{\rho_{11}}S_{12}\sqrt{\rho_{11}}& \sqrt{\rho_{11}}S_{12}
S_{12}^\dag\sqrt{\rho_{11}}+X_2^\dag
X_2\end{array}\right)\end{array}.$$ Thus $\rho$ is SPPT. Since
$[\rho^{T_A}]^{T_A}=\rho$, we get
$$\sqrt{\rho_{11}}S_{12}
S_{12}^\dag\sqrt{\rho_{11}}=\sqrt{\rho_{11}}S_{12}^\dag
S_{12}\sqrt{\rho_{11}}.
$$
This entails that $S_{12} S_{12}^\dag=S_{12}^\dag S_{12}$ as $\ker
S_{12}\supseteq \ker \sqrt{\rho_{11}}$, that is, $\rho$ is SSPPT.
\hfill$\blacksquare$

\textit{Example 1.} Let $\rho_{11}, D,T\in M_n({\mathbb C})$ with
$\rho_{11}\geq 0$, $\|D\|\leq 1$ and $\|T\|\leq 1$. Then the state
$\rho$ of the form
\begin{widetext}
\begin{eqnarray*}\rho=\frac{1}{{\rm
Tr}(\rho_{11}+\sqrt{\rho_{11}}DD^\dag
\sqrt{\rho_{11}})}\left(\begin{array}{cc}\rho_{11}&
\sqrt{\rho_{11}}T[\sqrt{\rho_{11}}DD^\dag
\sqrt{\rho_{11}}]^{\frac{1}{2}}\\
{[\sqrt{\rho_{11}}DD^\dag\sqrt{\rho_{11}}]^{\frac{1}{2}}T^\dag\sqrt{\rho_{11}}}&
\sqrt{\rho_{11}}DD^\dag \sqrt{\rho_{11}} \end{array}\right)
\end{eqnarray*}
\end{widetext}
in $M_2\otimes M_n$ is separable.

\section{Geometric measure of quantum discord}

The previous section showed that SSPPT is a sufficient but not
necessary condition of separability. In this section, in some sense
dual to it, we will prove that,  SSPPT is a necessary but not
sufficient condition of zero GMQD according to the very structure of
CQ states.

\subsection{Zero geometric measure of quantum discord states}

In order to propose a unified work for both finite- and
infinite-dimensional cases, we first
generalize QD and CQ to infinite-dimensional systems.

The quantum discord for infinite-dimensional systems was firstly
introduced and discussed in \cite{Datta}. For readers'
convenience, we restate this concept here.

{\it Quantum discord}\ \  Let $\rho\in\mathcal{S}(H_A\otimes H_B)$
with $\dim H_A\otimes H_B=+\infty$. Denote by
\begin{eqnarray*}
I(\rho):=S(\rho_A)+S(\rho_B)-S(\rho)
\end{eqnarray*}
 the quantum mutual information of $\rho$ whenever $S(\rho)<+\infty$, where
\begin{eqnarray*}
S(\rho):=-{\rm Tr}(\rho\log\rho)
\end{eqnarray*}
is the von Neumann entropy of the state $\rho$ (remark here that
$S(\rho)$ may be $+\infty$). Let $\Pi^A=\{\Pi_k^A=|k\rangle\langle
k|\}$ be a local von Neumann measurement. Performing $\Pi^A$ on
$\rho$, the outcome
\begin{eqnarray*}
\Pi^A(\rho)=\sum_kp_k\rho_k,
\end{eqnarray*}
where
\begin{eqnarray*}
\rho_k
=\frac{1}{p_k}(\Pi_k^A\otimes I_B)\rho(\Pi_k^A\otimes I_B)
\end{eqnarray*}
with probability
\begin{eqnarray*}
p_k={\rm Tr}[(\Pi_k^A\otimes I_B)\rho(\Pi_k^A\otimes I_B)].
\end{eqnarray*}
Define
\begin{eqnarray*}
I(\rho|\Pi^A):=S(\rho_B)-S(\rho|\Pi^A)
\end{eqnarray*}
and
\begin{eqnarray*}
S(\rho|\Pi^A):=\sum_kp_kS(\rho_k).
\end{eqnarray*}
If $I(\rho)<+\infty$, the difference
\begin{eqnarray}
D_A(\rho):=I(\rho)-\sup_{\Pi^A}I(\rho|\Pi^A)
\end{eqnarray}
is defined to be the quantum discord of $\rho$, where the supremum
is taken over all local von Neumann measurement.

$D_A(\rho)\geq0$ holds for any state
$\rho\in\mathcal{S}(H_A\otimes H_B)$ with $I(\rho)<+\infty$  since
the von Neumann entropy is strongly subadditive for both finite-
and infinite-dimensional cases (see \cite{Datta} for detail). One
can check that QD can also be calculated as
\begin{eqnarray}
D_A(\rho)=I(\rho)-\sup_{\Pi^A}I(\Pi^A(\rho)).
\end{eqnarray}
Namely, QD is defined as the infimum of
the difference of mutual information
of the pre-state $\rho$ and that of the
post-state $\Pi^A(\rho)$ with $\Pi^A$
runs over all local von Neumann measurements.

Symmetrically, one can define quantum discord $D_B$ with respect
to part B,  and the counterpart results are also valid. Note that
$D_A$ and $D_B$ are asymmetric, i.e.,
\begin{eqnarray*}
D_A(\rho)\neq D_B(\rho)
\end{eqnarray*}
in general.

 For
finite-dimensional systems, the classical-quantum (CQ) states
attracted much attention since they can be used for quantum
broadcasting \cite{Luo4}. It was point out in  \cite{Ollivier}
that  {\it a state is CQ if and only if it has zero quantum
discord}. Now we extend the concept of the CQ states to
infinite-dimensional case via the same scenario.

{\it Classical-quantum state}\ \  Similar to Eq.(\ref{y}), for
$\rho\in\mathcal{S}(H_A\otimes H_B)$, $\dim H_A\otimes
H_B=+\infty$, if $\rho$ admits a representation of the following
form
\begin{eqnarray}
\rho=\sum_kp_k|k\rangle\langle k|\otimes \rho_k^B,\label{b}
\end{eqnarray}
where $\{|k\rangle\}$ is an orthonormal set of $H_A$, $\rho_k^B$s
are states of the subsystem B, $p_k\geq0$ and $\sum_kp_k=1$, then we
call $\rho$ a classical-quantum (CQ) state.

However for infinite-dimensional case we do not know if every CQ
state has zero QD up to part A in general because for some CQ state
$\rho$ we may have $I(\rho)=+\infty$. So the concept of quantum
discord is not very suitable to the states in infinite-dimensional
systems. To get a more proper concept that can replace the concept
of quantum discord, we generalize the concept of the geometric
measure of quantum discord  \cite{Dakic} to infinite-dimensional
case.

Like to the finite-dimensional case, we define the geometric
measure of quantum discord up to part A of a state by
\begin{eqnarray}
D_A^G(\rho)=\inf \{\|\rho -\pi\|_2^2 : \pi \in
{\mathcal CQ}\},
\end{eqnarray}
where ${\mathcal CQ}$ is the set of all CQ states on $H_A\otimes
H_B$. That is, the geometric quantum discord of a state $\rho$ is
the square of the Hilbert-Schmidt distance of the state  to the
set of all CQ states. $D_A^G(\rho)$ makes sense for any state
$\rho$ because states are Hilbert-Schmidt operators.

It is known that, for a state $\rho$ in finite-dimensional system,
$D^G_A(\rho)=0$ if and only if $D_A(\rho)=0$, and in turn, if and
only if $\rho$ is CQ.

In the sequel we show that $D_A^G(\rho) =0$ if and only if $\rho$ is
a CQ state, and thus $D_A^G(\rho)$ is a suitable quantity replacing
$D_A(\rho)$. Before doing this let us firstly give a structural
feature of CQ states.

Write $\rho=\sum_{i,j}A_{ij}\otimes F_{ij}$ as in Eq.(\ref{z}).
For the case of $\dim H_A\otimes H_B<+\infty$, it is proved in
\cite{Ha} that, if $A_{ij}$s are mutually commuting normal
matrices, then $\rho$ is separable. We prove below that such a
state $\rho$ is not only separable but also a CQ state. In fact,
we give a characterization of CQ states in terms of commutativity
by  showing that $\rho$ is a CQ state if and only if $A_{ij}$s are
mutually commuting normal matrices. Moreover, this result is valid
for infinite-dimensional cases, too.

{\it Theorem 2.}\quad Let $\rho\in\mathcal{S}(H_A\otimes H_B)$ with
$\dim H_A\otimes H_B\leq+\infty$. Write $\rho=\sum_{ij}A_{ij}\otimes
F_{ij}$ with respect to some given bases of $H_A$ and $H_B$. Then
$\rho$ is a CQ state if and only if $A_{ij}$s are mutually commuting
normal operators acting on $H_A$.

{\it Proof.}\quad The `if' part. Assume that $A_{ij}$s are mutually
commuting normal operators, then $A_{ij}$s are simultaneously
diagonalizable since they are trace-class operators. Thus there
exist diagonal operators $D_{ij}$s and a unitary operator $U$ acting
on $H_A$ such that
\begin{eqnarray*}
(U^\dag\otimes I_B)\rho(U\otimes I_B)
=\sum_{i,j}D_{ij}\otimes F_{ij}.
\end{eqnarray*}
With no loss of generality, we may assume
\begin{eqnarray*}
\rho=\sum_{i,j}D_{ij}\otimes F_{ij}.
\end{eqnarray*}
It turns out that
$\rho$ can then be rewritten as
\begin{eqnarray*}
\rho=\sum_i \tilde{E}_{ii}\otimes B_{ii},
\end{eqnarray*}
where $\tilde{E}_{ii}$s are orthogonal rank-one  projections. Now it
is obvious that $B_{ii}\geq0$ since $\rho\geq0$, $i=1$, 2, $\dots$.
Hence, $\rho$ is a classical-quantum state.

The `only if' part.
If $\rho$ is a CQ state, then
\begin{eqnarray*}
\rho=\sum_kp_k|k\rangle\langle k|\otimes\rho_k^B,
\end{eqnarray*}
$p_k\geq0$, $\sum_kp_k=1$ for some orthonormal set $\{|k\rangle\}$
of $H_A$. Extend $\{|k\rangle\}$ to an orthonormal basis of $H_A$
and still denoted by $\{|k\rangle\}$. If $\Pi^A$ is a von Neumann
measurement induced from $\{|k\rangle\langle k|\}$, then it
follows from
\begin{eqnarray*}
\lefteqn{\quad\Pi^A(\rho)}\\
&&=\sum_k|k\rangle\langle k|
\otimes I_B(\sum_kp_k|k\rangle\langle k|
\otimes\rho_k^B)|k\rangle\langle k|\otimes I_B\\
&&=\rho
\end{eqnarray*}
that
\begin{align*}
&\sum_k|k\rangle\langle k|\otimes I_B
(\sum_{i,j}A_{ij}\otimes F_{ij})|k\rangle\langle k|\otimes I_B&\\
=&\sum_{i,j}A_{ij}\otimes F_{ij}.&
\end{align*}
This leads to
\begin{align*}
&\sum_k|k\rangle\langle k|A_{ij}|k\rangle\langle k|&\\
=&\sum_k\langle k|A_{ij}|k\rangle|k\rangle\langle k|
=A_{ij}&
\end{align*}
for any pair $(i,j)$, that is, every $A_{ij}$ is a diagonal operator
with respect to the same orthonormal basis $\{|k\rangle\}$.
Therefore, $A_{ij}$s are mutually commuting normal operators acting
on $H_A$. \hfill$\blacksquare$

Theorem 2 implies that CQ stems from noncommutativity but not from
entanglement. We can also find this kind of noncommutativity from
another perspective: for finite-dimensional case, it is proved in
\cite{Ferraro} that if $\rho$ is CQ (equivalently $D_A(\rho)=0$)
then
\begin{eqnarray*}
[\rho,\rho_A\otimes I_B]=0.
\end{eqnarray*}
It is easy to check that this result is  valid for
infinite-dimensional systems as well.

{\it Proposition 2.}\quad Let $\rho\in\mathcal{S}(H_A\otimes H_B)$
with $\dim H_A\otimes H_B\leq+\infty$. Then
\begin{eqnarray}
\rho\text{ is CQ } \Rightarrow[\rho,\rho_A\otimes I_B]=0.\label{x}
\end{eqnarray}

Indeed, if $\rho=\sum_{ij}A_{ij}\otimes F_{ij}$ as in Eq.(\ref{z})
with respect to some given bases of $H_A$ and $H_B$ and $\rho$ is a
CQ state, then $\rho_A=\sum_iA_{ii}$ commutes with $A_{ij}$ for any
$i$ and $j$. This ensures the commutativity of $\rho$ and
$\rho_A\otimes I_B$. So the noncommutativity signals quantumness of
the state. The converse is not true since for any state with maximal
marginal  we have Eq.(\ref{x}) holds in the finite-dimensional case
\cite{Ferraro}. One can check that the converse of Proposition 2 is
not true for infinite-dimensional case, either.

Theorem 2 is powerful for exploring the structure of CQ. For
instance, to prove the fact that CQ is equivalent to zero GMQD is
still valid for infinite-dimensional case, we need a geometric
feature of the set of all CQ states, that is, the set of all CQ
states is closed. This  can be proved by  applying Theorem 2.

{\sl Theorem 3.}\quad The set of all CQ states in ${\mathcal
S}(H_A\otimes H_B)$ is a closed set under  both the trace norm
topology and the Hilbert-Schmidt norm topology.

{\it Proof.}\quad Let $\rho$ be a state and
$\{\rho_n\}_{n=1}^\infty$ be a sequence of CQ states on $H_A\otimes
H_B$ such that $\lim_{n\rightarrow\infty}\rho_n=\rho$ under the
trace norm topology. For an arbitrarily chosen product basis
$\{|i\rangle|j^\prime\rangle\}_{i,j}$ of $H_A\otimes H_B$, $\rho_n$
and $\rho$ can be written in the form of
\begin{eqnarray*}
\rho_n=\sum_{i,j} A^{(n)}_{ij}\otimes F_{ij} \quad\mbox{and}\quad
\rho=\sum_{i,j} A_{ij}\otimes F_{ij},
\end{eqnarray*}
where $F_{ij}=|i^\prime\rangle\langle j^\prime|$ and $A_{ij}^{(n)},
A_{ij}\in {\mathcal T}(H_A)$. As $\rho_n$ is CQ, by Theorem 2,
$\{A_{ij}^{(n)}\}_{i,j}$ is a commutative set of normal operators
for each $n$. Now $\|A_{ij}^{(n)}-A_{ij}\|_{\rm Tr}=\|(I_A\otimes
|i^\prime\rangle\langle i^\prime|)(\rho_n-\rho)(I_A\otimes
|j^\prime\rangle\langle j^\prime|)\|_{\rm Tr}\leq
\|\rho_n-\rho\|_{\rm Tr}$ and $\lim_{n\rightarrow\infty}
\rho_n=\rho$ imply that $\lim_{n\rightarrow\infty}
A_{ij}^{(n)}=A_{ij}$ for each pair $(i,j)$. It follows that
$\{A_{ij}\}_{i,j}$ is a commutative set of normal operators, too,
which ensures that $\rho$ is a CQ  state by Theorem 2. Hence the set
of all CQ states is closed under the trace norm topology.

Since $\rho_n\rightarrow \rho$ under the Hilbert-Schmidt norm
topology if and only if $\rho_n\rightarrow \rho$ under the trace
norm topology for states $\rho_n$ and $\rho$, thus the set of all CQ
states is also closed under the Hilbert-Schmidt norm topology.
\hfill$\blacksquare$

Now the following theorem is obvious.

{\sl Theorem 4.}\quad Let $\rho\in{\mathcal S}(H_A\otimes H_B)$
with $\dim H_A\otimes H_B\leq \infty$ be a state. Then
$D^G_A(\rho)=0$ if and only if $\rho$ is a CQ state.

By Theorem 4 we get immediately the following known result for
finite-dimensional case.

{\sl Corollary 2.}\quad Let $\rho\in{\mathcal S}(H_A\otimes H_B)$
with $\dim H_A\otimes H_B< +\infty$ be a state. Then the following
statements are equiverlent.

(1) $ D_A(\rho)=0$.

(2) $D^G_A(\rho)=0$.

(3) $\rho$ is a CQ state.

 Symmetrically, we can define quantum-classical (QC)
states and it is clear that the counterpart results also valid.
Namely, we call $\rho$ a QC state if $\rho$ can be decomposed as
\begin{eqnarray}
\rho=\sum_jq_j\rho_j^A\otimes|j'\rangle\langle j'|,\label{c}
\end{eqnarray}
where $\{|j'\rangle\}$ is an orthonormal set of $H_B$, $\rho_j^A$s
are states of the subsystem A, $q_j\geq0$ and $\sum_jq_j=1$. We
can also define the geometric measure of quantum discord of a
state $\rho$ up to part B by
\begin{eqnarray}
D_B^G(\rho)=\inf \{\|\rho -\rho^\prime\|_2^2 : \rho^\prime \in
{\mathcal QC}\},
\end{eqnarray}
where ${\mathcal QC}$ is the set of all QC states on $H_A\otimes
H_B$.

The following results are obvious.

{\sl Theorem 2$^\prime$.}\quad Write $\rho=\sum_{k,l}F_{kl}\otimes
B_{kl}$ as in Eq.(\ref{a}). Then $\rho$ is a QC state if and only if
$B_{kl}$s are mutually commuting normal operators acting on $H_B$.

{\sl Proposition 2$^\prime$.}\quad Let
$\rho\in\mathcal{S}(H_A\otimes H_B)$ with $\dim H_A\otimes
H_B\leq+\infty$. Then $\rho$ is QC implies that $[\rho,I_A\otimes
\rho_B]=0$.

{\sl Theorem 3$^\prime$.}\quad The set of all QC states is closed.

{\sl Theorem 4$^\prime$.}\quad Let $\rho\in\mathcal{S}(H_A\otimes
H_B)$ with $\dim H_A\otimes H_B\leq+\infty$. Then $D_B^G(\rho)=0$
if and only if $\rho$ is a QC state.

\subsection{Witnessing (geometric measure of) quantum discord}

We now begin to discuss the relationship between zero GMQD states
and SSPPT states.

 \cite[Theorem 1]{Bylicka} claims that any classical-quantum (CQ)
state in $2\otimes n$ system with $n<\infty$ is not only SPPT up to
part A but also SPPT up to part B. And an example is given in
\cite{Bylicka} to illustrate that this conclusion is not valid for
$m\otimes n$ system if $m>2$.  However the example is not correctly
given. In fact, we remark here that the above conclusion is  valid
for any state in $m\otimes n$ with $m,n\leq +\infty$. Much more can
be achieved.

First observe that every CQ state is SSPPT up to A and every QC
state is SSPPT up to B. These can be checked directly by the
definitions.

The following main result of this subsection claims that a CQ/QC
state is not only SSPPT up to part A/B but also SSPPT up to part
B/A.

{\sl Theorem 5.}\quad Let $\rho\in\mathcal{S}(H_A\otimes H_B)$
with $\dim H_A\otimes H_B\leq+\infty$ be a state. If $\rho$ is QC
(CQ), then $\rho$ is SSPPT up to part B (up to part A). In
particular, if $\rho$ is not SSPPT up to part B or up to part A,
then both $D_B^G(\rho)$ and $D_A^G(\rho)$ are nonzero.

This result means that zero GMQD is much more stronger then SSPPT,
and any state with non-SSPPT  has quantum correlations tested by
GMQD.

{\sl Proof of Theorem 5.} \quad We only  check the former case
here, namely, the case  that $\rho$ is QC. We want to show that
$\rho$ is SSPPT up to part B. Write $\rho$ in the form
\begin{eqnarray*}
\rho=\sum_jq_j\rho_j^A\otimes|j'\rangle\langle j'|
\end{eqnarray*}
as in Eq.(\ref{c}). Then, by Theorem 2$^\prime$, $\rho$ can be
expressed as in Eq.(\ref{w}), where $B_{ij}$s are mutually commuting
normal trace-class  operators acting on $H_B$. We have to show that
there exists some $X$ of the form
\begin{eqnarray*}
X=\left(\begin{array}{c|c|c|c|c|c}
X_1&S_{12}X_1&S_{13}X_1&\cdots&S_{1m}X_1&\cdots \\ \hline
0&X_2&S_{23}X_2&\cdots&S_{2m}X_2&\cdots \\ \hline
\vdots&\vdots&\ddots&\vdots&\vdots&\cdots\\ \hline
0&0&0&X_{m-1}&S_{m-1,m}X_{m-1}&\cdots\\ \hline
0&0&0&0&X_m&\cdots\\ \hline
\vdots&\vdots&\vdots&\vdots&\vdots&\ddots
\end{array}\right)
\end{eqnarray*}
and
\begin{eqnarray*}
Y=\left(\begin{array}{c|c|c|c|c|c} X_1&S_{12}^\dag X_1&S_{13}^\dag
X_1&\cdots&S_{1m}^\dag X_1&\cdots \\ \hline 0&X_2&S_{23}^\dag
X_2&\cdots&S_{2m}^\dag X_2&\cdots \\ \hline
\vdots&\vdots&\ddots&\vdots&\vdots&\cdots\\ \hline
0&0&0&X_{m-1}&S_{m-1,m}^\dag X_{m-1}&\cdots\\ \hline 0&0&0&0&X_m&\cdots\\
\hline \vdots&\vdots&\vdots&\vdots&\vdots&\ddots
\end{array}\right)
\end{eqnarray*}
with $S_{ij}$s being diagonalizable operators satisfying
\begin{eqnarray*}
[S_{ki},S_{kj}^\dag]=0,\quad k<i\leq j,
\end{eqnarray*}
such that $\rho=X^\dag X$ and $\rho^{T_A}=Y^\dag Y$.

 \if false It is clear that there exists a unitary
operator $U$ on $H_B$ such that
\begin{eqnarray}
UB_{ij}U^\dag=D_{ij}
\end{eqnarray}
is a diagonal infinite matrix for any pair $(i, j)$, i.e., there is
an orthonormal basis $\{|n^{\prime\prime}\rangle\}$ of $H_B$ such
that
\begin{eqnarray*}
D_{ij}={\rm
diag}(b_1^{(ij)},b_2^{(ij)},\dots,b_n^{(ij)},\dots)=\sum_nb_n^{(ij)}
|n^{\prime\prime}\rangle\langle n^{\prime\prime}|
\end{eqnarray*}
for some family of complex numbers $\{b_n^{(ij)}\}$ satisfying
$\sum_{i,j,n}|b_n^{(ij)}|^2\leq 1$. Let
\begin{eqnarray*}
\Lambda(\cdot)=U(\cdot)U^\dag
\end{eqnarray*}
be a unitary channel of part B.
Then
\begin{eqnarray*}
I_A\otimes\Lambda(\rho)=[D_{ij}].
\end{eqnarray*}
It remains to show that $\tilde{\rho}=[D_{ij}]=\sum_{i,j}
E_{ij}\otimes D_{ij}$ is SSPPT up to part B.

\begin{eqnarray*}
\begin{array}{rl} \tilde{\rho}=&\sum_{i,j} E_{ij}\otimes (\sum_nb_n^{(ij)}
|n^{\prime\prime}\rangle\langle n^{\prime\prime}|)\\= &\sum_n
A_n\otimes G_{nn}=\left(\begin{array}{ccccc}
A_1&0&0&0&\cdots \\
0&A_2&0&0&\cdots \\
0&0&\ddots&0&\cdots\\
0&0&0&A_n&\cdots \\
\vdots&\vdots&\vdots&\vdots&\ddots
\end{array}\right),\end{array}
\end{eqnarray*}
where  $G_{nn}=|n^{\prime\prime}\rangle\langle n^{\prime\prime}| $
and $A_k=(b_k^{(ij)})_{ij}\geq 0$.\fi

Let $A_k=q_k\rho_k^A$. Then
\begin{eqnarray*}
{\rho}=\sum_k A_k\otimes F_{kk}=\left(\begin{array}{ccccc}
A_1&0&0&0&\cdots \\
0&A_2&0&0&\cdots \\
0&0&\ddots&0&\cdots\\
0&0&0&A_k&\cdots \\
\vdots&\vdots&\vdots&\vdots&\ddots
\end{array}\right),
\end{eqnarray*}
where  $F_{kk}=|k^{\prime}\rangle\langle k^{\prime}| $.

Thus there are upper triangular Hilbert-Schmidt operators $Z_k$ on
$H_A$ such that
\begin{eqnarray*}
A_k=Z_k^\dag Z_k,\quad k=1, 2, \dots,
\end{eqnarray*}
and then
\begin{eqnarray*}
{\rho}=Z^\dag Z
\end{eqnarray*}
with
\begin{eqnarray*}
Z=\sum_k Z_k\otimes F_{kk}.
\end{eqnarray*}
Write
\begin{eqnarray*}
Z_k=(z_{ij}^{(k)})_{i,j}
\end{eqnarray*}
with $z_{ij}^{(k)}=0 $ whenever $i<j$ and let
\begin{eqnarray*}
 X=(X_{ij})_{i,j},
 \end{eqnarray*}
where
\begin{eqnarray*}
X_{ij}={\rm diag}(z_{ij}^{(1)}, z_{ij}^{(2)}, \cdots, z_{ij}^{(n)},
\cdots)=\sum_n z_{ij}^{(n)}|n^\prime\rangle\langle n^\prime |.
\end{eqnarray*}
Then $X_{ij}=0$ if $i<j$ and
\begin{eqnarray*}
Z=X=\sum_{i,j} E_{ij}\otimes X_{ij},
\end{eqnarray*}
which  is   an upper triangular operator matrix and ${\rho}=X^\dag
X$.

By  Lemma 1 below, we can choose $Z_k$s so that $z_{ii}^{(n)}=0$
implies that $z_{ij}^{(n)}=0$. It turns out  $X_{ij}$ can be
written in $X_{ij}=S_{ij}X_{ii}$ for some diagonal operator
$S_{ij}=\sum_n s_{ij}^{(n)}|n^\prime\rangle\langle n^\prime |$ for
any $(i,j)$ with $i<j$. Obviously we have $[S_{ij},
S_{il}^\dag]=0$ for any $i<j\leq l$.

Now it is easily checked that $\rho^{T_A}=Y^\dag Y$. Hence $\rho$ is
SSPPT up to part B.

Finally, $D_A^G(\rho)=0$ if and only if $\rho$ is CQ. By what
proved above, $\rho$ then is SSPPT up to A as well as up to B. So,
$\rho$ is not SSPPT up to A or up to B will implies that both
$D_A^G(\rho)>0$ and $D_B^G(\rho)>0$ hold. \hfill$\blacksquare$

{\sl Lemma 1}\quad Let $A$ be a positive finite or infinite matrix
with $A=X^\dag X$ for some finite or infinite upper triangular
matrix $X=(x_{ij})_{i,j}$, $\|A\|_{\rm Tr}<+\infty$. If $x_{kk}=0$
for some $k$, then there exists a finite or infinite upper
triangular matrix $Y=(y_{ij})_{i,j}$ with $y_{kj}=0$, $j=1$, 2,
$\dots$, such that $A=Y^\dag Y$.

{\sl Proof.}\quad Let
\begin{eqnarray*}
X=\left(\begin{array}{cccccc}
x_{11}&x_{12}&x_{13} &\cdots&x_{1n}&\cdots \\
&{x}_{22}&x_{23}&\cdots&{x}_{2n}&\cdots \\
&&x_{33}&\cdots&x_{3n}&\cdots\\
&&&\ddots&\vdots&\vdots\\
&&&&x_{nn}&\cdots\\
&&&&&\ddots
\end{array}\right),
\end{eqnarray*}
\begin{eqnarray*}
|\eta_1\rangle=\left(\begin{array}{c}
\bar{x}_{11} \\
\bar{x}_{12} \\
\vdots\\
\bar{{x}}_{1n}\\
\vdots
\end{array}\right),
|\eta_2\rangle=\left(\begin{array}{c}
0 \\
\bar{x}_{22} \\
\bar{x}_{23} \\
\vdots\\
\bar{{x}}_{2n}\\
\vdots
\end{array}\right),\dots,
\end{eqnarray*}
\begin{eqnarray*}
\begin{array}{c@{\hspace{-5pt}}l}
|\eta_k\rangle=\left(\begin{array}{c}
0 \\
\vdots \\
0\\
\bar{{x}}_{kk}\\
\vdots \\
\bar{{x}}_{k,k+1}\\
\bar{{x}}_{k,k+2}\\
\vdots
\end{array}\right)
&\begin{array}{l}\left.\rule{0mm}{7mm}\right\}k-1 \\
\\ \\ \\ \\ \\ \\  \end{array}
\end{array},\dots.
\end{eqnarray*}
Then
\begin{eqnarray}
A=\sum_i|\eta_i\rangle\langle\eta_i|.
\end{eqnarray}
If $x_{kk}=0$, we let
\begin{eqnarray}
A=\sum_{i=1}^{k-1}|\eta_i\rangle\langle\eta_i|
+0\oplus A_{k+1},
\end{eqnarray}
where $0$ is a $k\times k$ zero matrix.
Let
\begin{eqnarray*}
A_{k+1}=\widetilde{X}^\dag_{k+1} \widetilde{X}_{k+1}
\end{eqnarray*}
with
\begin{widetext}
\begin{eqnarray*}
\widetilde{X}_{k+1}=\left(\begin{array}{cccccc}
x'_{11}&x'_{12}&x'_{13} &\cdots&x'_{1n}&\cdots \\
&{x'}_{22}&x'_{23}&\cdots&{x'}_{2n}&\cdots \\
&&x'_{33}&\cdots&x'_{3n}&\cdots\\
&&&\ddots&\vdots&\vdots\\
&&&&x'_{nn}&\cdots\\
&&&&&\ddots
\end{array}\right).
\end{eqnarray*}

Taking \vspace{5mm}

\begin{eqnarray}
Y=\left(\begin{array}{ccccccccc}
x_{11}&x_{12}&\cdots &x_{1,k-1}&x_{1k}&x_{1,k+1}&\cdots&x_{1n}&\cdots \\
&x_{22}&\cdots &x_{2,k-1}&x_{2k}&x_{2,k+1}&\cdots&x_{2n}&\cdots \\
&&\ddots&\vdots&\vdots&\vdots&\vdots&\vdots&\vdots\\
&&&x_{k-1,k-1}&x_{k-1,k}&x_{k-1,k+1}&\cdots&x_{k-1,n}&\cdots \\
&&&&0&0&\cdots&0&\cdots \\
&&&&&x'_{11}&\cdots&x'_{1,n-k}&\cdots\\
&&&&&&\ddots&\vdots&\vdots\\
&&&&&&&x'_{n-k,n-k}&\cdots\\
&&&&&&&&\ddots
\end{array}\right),
\end{eqnarray}
\end{widetext}
it is clear that $A=Y^\dag Y$ as desired.\hfill$\blacksquare$

\if false
 \cite[Theorem 1]{Bylicka} claims that any classical-quantum (CQ)
state in $2\otimes n$ system is not only SPPT up to part A but also
SPPT up to part B. We remark here that this fact is still valid for
the case $n=+\infty$. Reviewing the proof of \cite[Theorem
1]{Bylicka}, we can get that the results above are still true if we
replace ``SPPT'' by ``SSPPT''. In fact, if $\rho$ is a CQ state
acting on $H_A\otimes H_B$ with $\dim H_A=2$ and $\dim
H_B\leq+\infty$, then
\begin{eqnarray}
\rho=\sum\limits_{i=1}^2p_i|\psi_i\rangle\langle\psi_i|\otimes\sigma_i, \label{q}
\end{eqnarray}
where $\{|\psi_1\rangle, |\psi_2\rangle\}$ is a basis in $\mathbb{C}^2$, $\sigma_i$s are states on $H_B$.
Let $U=\left(\begin{array}{cc}u_{11}&u_{12}\\
u_{21}&u_{22}\end{array}\right)$ be a $2\times 2$ unitary matrix and let $U|\psi_i\rangle=|e_i\rangle$.
The block structure of Eq.(\ref{q}) in the canonical computational basis $\{|e_1\rangle, |e_2\rangle\}$
reads as:
\begin{eqnarray*}
\rho=\left(\begin{array}{cc}\rho_{11}&\rho_{12}\\
\rho_{12}^\dag&\rho_{22}\end{array}\right),
\end{eqnarray*}
where
\begin{eqnarray*}
&\rho_{11}=p_1|u_{11}|^2\sigma_1+p_2|u_{12}|^2\sigma_2&\\
&\rho_{22}=p_1|u_{21}|^2\sigma_1+p_2|u_{22}|^2\sigma_2,&\\
&\rho_{12}=p_1u_{11}\bar{u}_{21}\sigma_1+p_2u_{12}\bar{u}_{22}\sigma_2.&
\end{eqnarray*}
One has, therefore,
\begin{eqnarray*}
X_1^dag X_1=p_1|u_{11}|^2\sigma_1
+p_2|u_{12}|^2\sigma_2,
\end{eqnarray*}
and hence one may take
\begin{eqnarray*}
X_1=(p_1|u_{11}|^2\sigma_1
+p_2|u_{12}|^2\sigma_2)^{\frac{1}{2}}.
\end{eqnarray*}
Clearly, $X_1$ is positive.
Then we have
\begin{eqnarray*}
X_1^\dag SX_1=p_1u_{11}\bar{u}_{21}\sigma_1+p_2u_{12}\bar{u}_{22}\sigma_2,
\end{eqnarray*}
which gives rise to
\begin{eqnarray*}
S=X_1^{-1}(p_1u_{11}\bar{u}_{21}\sigma_1+p_2u_{12}\bar{u}_{22}\sigma_2)X_1^{-1}.
\end{eqnarray*}
Here $X_1^{-1}:=\sum\limits_k\lambda_k^{-1}|\phi_k\rangle\langle\phi_k|$
whenever $X_1=\sum\limits_k\lambda_k|\phi_k\rangle\langle\phi_k|$.
Now, taking into account
\begin{eqnarray*}
u_{11}\bar{u}_{21}+u_{12}\bar{u}_{22}=0
,\end{eqnarray*}
one obtains
\begin{eqnarray*}
S=u_{11}\bar{u}_{21}X_1^{-1}(p_1\sigma_1-p_2\sigma_2)X_1^{-1}.
\end{eqnarray*}
Notice that $X_1^{-1}(p_1\sigma_1-p_2\sigma_2)X_1^{-1}$ is Hermitian, hence
$S$ is normal which implies that $\rho$ is SSPPT up to part B.
It is also showed in \cite{Bylicka} that for $m\otimes
n$ systems with $m>2$, CQ does not guarantee SPPT up to part B.
Symmetrically, these results are also true for $n\otimes 2$ systems:

{\sl Proposition 3.}\quad Assume that $\dim H_B=2$, $\dim
H_A=n\leq+\infty$ and $\rho\in\mathcal{S}(H_A\otimes H_B)$. If
$\rho$ is QC, then $\rho$ is a SSPPT state up to part A.

That is, together with Theorem 3, we get that if a state in
$n\otimes 2$ (resp. $2\otimes n$)systems is neither SSPPT up to part
A nor SSPPT up to part B, then $\rho$ contains quantum correlations
measured by QD with respect to  part B (resp. A). However, this
result does not hold for $\dim H_B>2$ (resp. $\dim H_A>2$) according
to \cite{Bylicka}. \fi

Reviewing the discussion above, we also know that SSPPT up to part
B/A is only a necessary condition of zero GMQD.

\textit{Example 2.}\quad The so-called circulant state in $2\otimes
2$ system \cite{Chruscinski2} is given by
\begin{eqnarray*}
\rho=\left(\begin{array}{cc|cc}
a_{11}&0&0&a_{12} \\
0&b_{11}&b_{12}&0 \\ \hline
0&b_{21}&b_{22}&0\\
a_{21}&0&0&a_{22}\end{array}\right).
\end{eqnarray*}
Assume that $a_{11}b_{11}>0$. It can be derived form
\cite{Chruscinski} that $\rho$ is SSPPT up to part B if and only if
$\tilde{a}\geq0$, $\tilde{b}\geq0$ and $|a_{12}|=|b_{12}|$, where
\begin{eqnarray*}
\tilde{a}=\left(\begin{array}{cc}
a_{11}&b_{21} \\
b_{12}&a_{22}\end{array}\right),\quad
\tilde{b}=\left(\begin{array}{cc}
b_{11}&a_{21} \\
a_{12}&b_{22}\end{array}\right).
\end{eqnarray*}
On the other hand it is easily checked that $\rho$ is a QC state if
and only if $a_{11}=b_{11}$, $a_{22}=b_{22}$ and
$|a_{12}|=|b_{12}|$, which implies that there exist SSPPT states
that are not QC states.

Facts listed above clearly indicate that zero GMQD (equivalently,
zero QD for finite-dimensional case) has strong local
commutativity than that of SSPPT.

\section{Nullity of measurement-induced nonlocality}

In Secs.II-III we comparing SSPPT with separability, CQ, QC and zero
QD(GMQD) respectively by means of local commutativity. The present section
is devoted to the nullity of MiN (Measurement-induced nonlocality)
in terms of local commutativity, from which we get a clearer picture
of these different quantum correlations.

With the same spirit as that of the finite-dimensional case, we
first generalize the concept of MiN to infinite-dimensional
bipartite systems.

{\it Measurement-induced nonlocality}\ \ Assume that $\dim
H_A\otimes H_B=+\infty$ and $\rho\in\mathcal{S}(H_A\otimes H_B)$.
Let $\Pi^A=\{\Pi_k^A=|k\rangle\langle k|\}$ be a set of mutually
orthogonal rank-one projections that sum up to the identity of
$H_A$. Similar to the finite-dimensional case, we call such
$\Pi^A=\{\Pi_k^A\}$ a local von Neumann measurement. Note that
$\sum_k(\Pi_k^A\otimes I_B)^\dag(\Pi_k^A\otimes I_B)
=\sum_k\Pi_k^A\otimes I_B=I_{AB}$, here the series converges under
the strongly operator topology \cite{Hou}. We define the
Measurement-induced nonlocality (MiN, briefly) of $\rho$ by
\begin{eqnarray}
N_A(\rho):=\sup_{\Pi^A}\|\rho-\Pi^A(\rho)\|_2^2,
\end{eqnarray}
where the supremum is taken over all local von Neumann measurement
$\Pi^A=\{\Pi_k^A\}$ that satisfying
\begin{eqnarray}
\sum_k\Pi_k^A\rho_A\Pi_k^A=\rho_A.
\end{eqnarray}

The following properties are straightforward.

(i) $N_A(\rho)=0$ for any product
state $\rho=\rho_A\otimes \rho_B$.

(ii) $N_A(\rho)$ is locally unitary invariant, namely,
$N_A((U\otimes V)\rho(U^\dag\otimes V^\dag))=N_A(\rho)$ for any
unitary operators $U$ and $V$ acting on $H_A$ and $H_B$,
respectively.

(iii) $N_A(\rho)>0$ whenever $\rho$ is
entangled since $\Pi^A(\rho)$ is
always a classical-quantum state and thus is separable.

(iv) $0\leq N_A(\rho)<4$.

The MiN of a pure state can be easily calculated. Let
$|\psi\rangle\in H_A\otimes H_B$ and
\begin{eqnarray*}
|\psi\rangle=\sum_k\lambda_k|k\rangle|k'\rangle
\end{eqnarray*}
be its Schmidt decomposition. For the
finite-dimensional case,
Luo and Fu showed in \cite{Luo1} that
\begin{eqnarray}
N_A(|\psi\rangle)= 1-\sum_k\lambda_k^4.
\end{eqnarray}
 This is also true for pure states in
infinite-dimensional systems. Dually, one can define MiN with
respect to the second subsystem B---$N_B$, and the corresponding
properties are valid. It is easily seen that these two MiNs are
asymmetric, namely, the MiN with respect to subsystem A is not equal
to the one with respect to subsystem B generally.

Let us now begin to discuss the nullity of MiN. The following is the
main result of this section (we only discuss the case of $N_A$ since
the case of $N_B$ can be obtained by interchanging the role of A and
B).

{\it Theorem 6.}\quad Let $\rho\in\mathcal{S}(H_A\otimes H_B)$ be a
state with $\dim H_A\otimes H_B\leq+\infty$. Let $\{|k\rangle\}$ and
$\{|i'\rangle\}$ be any orthonormal bases of $H_A$ and $H_B$,
respectively. Write $\rho=\sum_{i,j}A_{ij}\otimes F_{ij}$ as in
Eq.(\ref{z}) with respect to the given bases. Then $N_A(\rho)=0$ if
and only if $A_{ij}$s are mutually commuting normal operators and
each eigenspace of $\rho_A$ contained in some eigenspace of $A_{ij}$
for all $i$ and $j$.

{\it Proof.}\quad By the definition
of $N_A(\rho)$, it is clear that the condition
$N_A(\rho)=0$ is equivalent to the condition
that $\Pi^A(\rho)=\rho$ holds for any local von Neumann
measurement that make $\rho_A$ invariant.

The `if' part. If each eigenspace of $\rho_A$
is a one-dimensional space,
then $\rho_A=\sum_ip_i|i\rangle\langle i|$
for some orthonormal base $\{|i\rangle\}$
and $\{p_i\}$ with $p_i>0$, $p_i\neq p_j$ if $i\neq j$.
Obviously, for any local von Neumann
measurement $\Pi^A=\{\Pi_k^A\}$,
$\sum_k\Pi_k^A\rho_A\Pi_k^A=\rho_A$ implies that,
for each $k$, $|k\rangle=|i\rangle$ for
some $i$. Thus $\Pi^A$ is introduced in
fact by $\{|i\rangle\}$.
Now it is clear that $\Pi^A(\rho)=\rho$
as every $A_{ij}$ commutes with
$\rho_A$.

Denote
\begin{eqnarray*}
E(\lambda^A)={\ker}(\lambda^A-\rho_A)
\end{eqnarray*}
(here, ${\ker}(X)$ stands for the kernel of the operator $X$), and
assume that $\dim {\ker}(\lambda^A-\rho_A)\geq2$ for some nonzero
eigenvalue $\lambda^A$ of $\rho_A$. Then the restricted operator of
$\rho_A$ on $E(\lambda^A)$, denoted by $\rho_A|_{E(\lambda^A)}$,
satisfying
\begin{eqnarray*}
\rho_A|_{E(\lambda^A)}=\lambda^A I_{E(\lambda^A)},
\end{eqnarray*}
where $I_{E(\lambda^A)}$ is the identity operator on $E(\lambda^A)$.
As $A_{ij}$s are mutually commuting normal operators and each
eigenspace of $\rho_A$ contained in some eigenspace of $A_{ij}$ for
all $i$ and $j$, we see that
\begin{eqnarray*}
C_{ij}=A_{ij}|_{E(\lambda^A)}=\lambda^{(ij)} I_{E(\lambda^A)}
\end{eqnarray*}
for some eigenvalue
$\lambda^{(ij)}$ of
$A_{ij}$ for any $i$ and $j$. This leads to
\begin{eqnarray*}
\sum_k\Pi_k^A A_{ij}\Pi_k^A=A_{ij}
\end{eqnarray*}
for any local von Neumann measurement
$\Pi^A=\{\Pi_k^A\}$ that doesn't disturb $\rho_A$ locally, so
we have $\Pi^A(\rho)=\rho$.
\if The sufficiency is now concluded.\fi

The `only if' part. If $\Pi^A(\rho)=\rho$ for
any local von Neumann measurement $\Pi^A$ that leave $\rho_A$
invariant, then $\Pi^A$ satisfying
\begin{eqnarray*}
\sum_k\Pi_k^A A_{ij}\Pi_k^A=A_{ij}
\end{eqnarray*}
for any $i$, $j$.
This forces that $A_{ij}$s are
mutually commuting normal operators.
We show that
each eigenspace of $\rho_A$ contained in some eigenspace of
$A_{ij}$ for all $i$ and $j$.
Or else, we may assume with no loss of generality that
\begin{eqnarray*}
\dim {\ker}(\lambda^{(i_0j_0)}-A_{i_0j_0})=1
\end{eqnarray*}
while
\begin{eqnarray*}
\dim {\ker}(\lambda^{A}-\rho_A)=2
\end{eqnarray*}
for some nonzero eigenvalue $\lambda^{(i_0j_0)}$ of $A_{i_0j_0}$ and
nonzero eigenvalue $\lambda^A$ of $\rho_A$. It turns out that there
must exist an orthonormal basis of ${\ker}(\lambda^{A}-\rho_A)$,
denoted by $\{|e_1\rangle$, $|e_2\rangle\}$, and a local von Neumann
measurement $\Pi^A$ induced from an orthonormal basis containing
$\{|e_1\rangle$, $|e_2\rangle\}$ such that
$\sum_k\Pi^A(\rho_A)\Pi_k^A=\rho_A$ while
$\sum_k\Pi_k^AA_{i_0j_0}\Pi_k^A\not=A_{i_0j_0}$, a contradiction.
\hfill$\blacksquare$

Symmetrically, we have

{\it Theorem 6$^\prime$.}\quad Let
$\rho=\sum_{k,l}F_{kl}\otimes B_{kl}$
as in Eq.~(\ref{a})
with respect to the given bases.
Then $N_B(\rho)=0$ if and only if $B_{kl}$s are mutually
commuting normal operators and
each eigenspace of $\rho_B$ contained in some eigenspace of
$B_{kl}$ for all $k$ and $l$.

Theorem 6 and 6$^\prime$ indicate that the phenomenon of MiN is
a manifestation of quantum correlations due to
noncommutativity rather than due to entanglement as well.
And we claim that the
commutativity for a state to have zero MiN is
`stronger' than that of zero QD(GMQD) state.
We illustrate it with the following example.

{\it Example 3.}\quad We consider a $3\otimes 2$ system.
Let
\begin{eqnarray*}
\rho=\left(\begin{array}{ccc|ccc}
a&0&0&e&0&0\\
0&a&0&0&f&0\\
0&0&b&0&0&g \\ \hline
\bar{e}&0&0&c&0&0 \\
0&\bar{f}&0&0&c&0\\
0&0&\bar{g}&0&0&d \\
\end{array}\right).
\end{eqnarray*}
It is clear that $\rho$ is a CQ state for any positive
numbers $a$, $b$, $c$, $d$ and complex numbers $e$, $f$, $g$ that
make $\rho$ be a state. However,
taking $\Pi^A=\{|\psi_i\rangle\langle\psi_i|\}_{i=1}^3$
with
\begin{eqnarray*}
|\psi_1\rangle=\frac{1}{\sqrt{2}}\left(\begin{array}{c}1\\1\\0\end{array}\right),
|\psi_2\rangle=\frac{1}{\sqrt{2}}\left(\begin{array}{c}1\\-1\\0\end{array}\right),
|\psi_3\rangle=\left(\begin{array}{c}0\\0\\1\end{array}\right),
\end{eqnarray*}
it is easy to see that
\begin{eqnarray*}
\sum_k\Pi_k^A\rho_A\Pi_k^A=\rho_A
\end{eqnarray*}
and
\begin{eqnarray*}
\Pi^A(\rho)\neq\rho \text{ whenever } e\neq f.
\end{eqnarray*}
If $a+c=b+d$, it is easily checked   that $N_A(\rho)=0$ if and only
if $a=b$, $c=d$ and $e=f=g$. Hence, there are many CQ states with
nonzero MiN.

Let
\begin{eqnarray*}
&\mathcal{S}_{N_{A/B}}^0
=\{\rho\in\mathcal{S}(H_A\otimes H_B): N_{A/B}(\rho)=0\},&
\end{eqnarray*}
\begin{eqnarray*}
&\mathcal{S}_{D_{A/B}^G}^0
=\{\rho\in\mathcal{S}(H_A\otimes H_B): D_{A/B}^G(\rho)=0\},&
\end{eqnarray*}
\begin{eqnarray*}
&\mathcal{CQ}
=\{\rho\in\mathcal{S}(H_A\otimes H_B): \rho\ \mbox{\rm is CQ}\},&
\end{eqnarray*}
\begin{eqnarray*}
&\mathcal{QC}
=\{\rho\in\mathcal{S}(H_A\otimes H_B): \rho\ \mbox{\rm is QC}\}&
\end{eqnarray*}
and $\mathcal{S}_{sep}$ be the set of all
separable states acting on $H_A\otimes H_B$.
The above example shows that, $\mathcal{S}_{N_{A/B}}^0$
is a proper subset of $\mathcal{CQ}/\mathcal{QC}$.
In addition, for $0\leq\epsilon\leq1$, $\rho_1$,
$\rho_2\in\mathcal{S}_{N_{A/B}}^0$
do not imply
$\epsilon\rho_1+(1-\epsilon)\rho_2\in\mathcal{S}_{N_{A/B}}^0$ in
general, so $\mathcal{S}_{N_{A/B}}^0$ is not a
convex set. Similarly, $\mathcal{S}_{D_{A/B}^G}^0$,
$\mathcal{CQ}$ and $\mathcal{QC}$ are not
convex, either.

Furthermore, equivalent to Theorem 6 and 6$^\prime$, one can check that

{\sl Corollary 3.}\quad Let $\rho\in\mathcal{S}(H_A\otimes H_B)$
with $\dim H_A\otimes H_B\leq+\infty$ be a state. Then

(i) $N_{A}(\rho)=0$ if and only if
\begin{eqnarray*}
\rho=\sum_kp_k|k\rangle\langle k|\otimes\rho_k^B
\end{eqnarray*}
with $\rho_k^B=\rho_l^B$ whenever $p_k=p_l$;

(ii) $N_{B}(\rho)=0$ if and only if
\begin{eqnarray*}
\rho=\sum_jq_j\rho_j^A\otimes|j'\rangle\langle j'|
\end{eqnarray*}
with $\rho_j^A=\rho_i^A$ whenever $q_j=q_i$.

Comparing with Eqs.~(\ref{b}) and (\ref{c}),
we get a more transparent picture of these two
different quantum correlations.

Reviewing the proof of Theorem 6 and 6$^\prime$, the following is
clear:

{\it Proposition 4.}\quad  Let $\rho\in\mathcal{S}(H_A\otimes H_B)$,
$\dim H_A\otimes H_B\leq+\infty$. Suppose that each eigenspace of
$\rho_A$ (resp. $\rho_B$) is of one-dimension and
$\rho_A=\sum_kp_k|k\rangle\langle k|$ (resp.
$\rho_B=\sum_lq_l|l'\rangle\langle l'|$) is the spectral
decomposition. Then the local von Neumann measurement $\Pi^A$ (resp.
$\Pi^B$) that makes $\rho_A$ (resp. $\rho_B$) invariant is uniquely
(\emph{up to permutation}) induced from $\{|k\rangle\langle k|\}$
(resp. $\{l'\rangle\langle l'|\}$), and vice versa.

In Ref.\cite{Luo1}, for finite-dimensional case, the authors claim
that $N_A(\rho)$ =0 for any classical-quantum state
$\rho=\sum_kp_k|k\rangle\langle k|\otimes\rho_k^b$ whose marginal
state $\rho^a=\sum_kp_k|k\rangle\langle k|$ is nondegenerate (here,
a matrix $A$ is said to be nondegenerate provided that each
eigenspace of $A$ is of one-dimension). This is also valid for
infinite-dimensional case.

{\it Corollary 4.}\quad Assume that  $\rho\in\mathcal{S}(H_A\otimes
H_B)$ with $\dim H_A\otimes H_B\leq+\infty$. If
$\rho\in\mathcal{CQ}$ (resp. $\mathcal{QC}$), then $N_A(\rho)=0$
(resp. $N_B(\rho)=0$) provided that each eigenspace of $\rho_A$
(resp. $\rho_B$) is of one-dimension.

It is known that, for the finite-dimensional case,
$\mathcal{S}_{D_{A/B}^G}^0$ is a zero-measure set \cite{Ferraro}
(that is, each point of this set can be approximated by a sequence
of states that not belong to this set with respect to the trace
norm), and, for the infinite-dimensional case, $\mathcal{S}_{sep}$
is a zero-measure set \cite{Clifton}. Thus,
both$\mathcal{S}_{N_{A/B}}^0$ and $\mathcal{S}_{D_{A/B}^G}^0$ are
zero-measure set in both finite- and infinite-dimensional cases.
This indicates that MiN is ubiquitous: almost all quantum states
have nonzero MiN. In other words, as a resource, we get more states
valid in tasks of quantum processing based on MiN.

\section{Conclusions}

In terms of local commutativity, for both finite- and
infinite-dimensional systems, we show that (1) SSPPT states are
countably separable, (2) SSPPT can detects QD(GMQD), and
furthermore, the zero MiN states and the zero GMQD states are
characterized. We argue that MiN is the most essential quantum
correlation among MiN, QD, GMQD and entanglement. They all
originated from the \emph{supposition} of the states (since for a
pure state $\rho$, it is separable if and only if
$N_{A/B}(\rho)=D_{A/B}^G(\rho)=0$).

As a result, we obtain the following chain of (proper) inclusions
for finite-dimensional case:
\begin{eqnarray}\begin{array}{rl}
\mathcal{S}_p\subset&\mathcal{S}_{N_{A/B}}^0\subset\mathcal{CQ/QC}={\mathcal
S}^0_{D_{A/B}} =\mathcal{S}_{D_{A/B}^G}^0\\
\subset&\mathcal{S}_{SSPPT}^{A}\cap\mathcal{S}_{SSPPT}^{B}
\subset\mathcal{S}_{SSPPT}^{A/B} \\\subset&\mathcal{S}_{csep}
=\mathcal{S}_{sep}\subset \mathcal{PPT},\end{array}
\end{eqnarray}
and chain of (proper) inclusions for infinite-dimensional case:
\begin{eqnarray}\begin{array}{rl}
\mathcal{S}_p\subset&\mathcal{S}_{N_{A/B}}^0\subset\mathcal{CQ/QC} =\mathcal{S}_{D_{A/B}^G}^0\\
\subset&\mathcal{S}_{SSPPT}^{A}\cap\mathcal{S}_{SSPPT}^{B}
\subset\mathcal{S}_{SSPPT}^{A/B} \\\subset&\mathcal{S}_{csep}
\subset\mathcal{S}_{sep}\subset \mathcal{PPT},\end{array}
\end{eqnarray}
where $\mathcal{S}_p$ denotes the set of all product states,
$S^0_{D_{A/B}}$ is the set of all zero QD states, i.e.,
$S^0_{D_{A/B}}=\{ \rho\in{\mathcal S}(H_A\otimes H_B):
D_{A/B}(\rho)=0\}$, $\mathcal{S}_{SSPPT}^{A/B}$ denotes the set of
all SSPPT states up to part A/B, $\mathcal{S}_{csep}$ stands for the
set of all countably separable states and $\mathcal{S}_{sep}$
denotes the set consisting of all separable states and
$\mathcal{PPT}$ denotes the set of all PPT states.

The above inclusion chains indicate that the weaker quantum
correlation is, the stronger commutativity is. Consequently, we may
guess   that $\rho$ is separable if and only if its local operators
$A_{ij}$s or $B_{kl}$s have certain ``commutativity'' properties of
some degree. This is an interesting task and worth to make a further
research.

In addition, our results also suggest several questions for further
studies such as, (i) comparing $N_{A/B}(\rho)$ with
$D_{A/B}^G(\rho)$ and some other entanglement measures (such as
concurrence or entanglement of formation), and (ii) establishing
computable formula of $N_{A/B}(\rho)$ for   arbitrary state $\rho$
for both finite- and infinite-dimensional cases.

\begin{acknowledgments}
This work is partially supported by Natural Science Foundation of
China (11171249, 11101250), Research Fund for the Doctoral Program
of Higher Education of China (20101402110012) and Research start-up
fund for the Doctors of Shanxi Datong University.
\end{acknowledgments}

\nocite{*}

\bibliography{apssamp}

\end{document}